\documentclass[referee]{aa}
\usepackage{amsmath}
\usepackage{graphicx}
\usepackage{txfonts}
\usepackage{natbib}
\usepackage{multirow}

\bibpunct{(}{)}{;}{a}{}{,} 

\begin{document}

\title{Short-term forecast of the total and spectral solar irradiance}

\author{Luis Eduardo A. Vieira\inst{1}, Thierry Dudok de Wit \inst{1},  Matthieu Kretzschmar \inst{1,2}}

\offprints{L.E.A. Vieira, \email{luis.vieira@cnrs-orleans.fr}}

\institute{Laboratoire de Physique et Chimie de l'Environnement et de l'Espace  (LPC2E/CNRS), 3A, Avenue de la Recherche, 45071 Orléans cedex 2, France \and Observatoire Royal de Belgique, Avenue Circulaire, 3, B-1180 Brussels, Belgium}

\date{Received  / Accepted }

\abstract {The conditions of the upper atmosphere can change rapidly in response to the solar and geomagnetic activity. Among several heliophysical and geophysical quantities, the accurate evolution of the solar irradiance is fundamental to forecast the evolution of the neutral and ionized components of the Earth's atmosphere.}{We developed an artificial neural network model to compute the evolution of the solar irradiance in near-real time. The model is based on the assumption that that great part of the solar irradiance variability is due to the evolution of the structure of the solar magnetic field.}{We employ a Layer-Recurrent Network (LRN) to model the complex relationships between the evolution of the bipolar magnetic structures (input) and the solar irradiance (output). The evolution of the bipolar magnetic structures is obtained from near-real time solar disk magnetograms and intensity images. The magnetic structures are identify and classified according to the area of the solar disk covered. We constrained the model by comparing the output of the model and observations of the solar irradiance made by instruments onboard of SORCE spacecraft. Here we focus on two regions of the spectra that are covered by SORCE instruments. While the range from 115 nm to 310 nm is covered by the two SOLSTICE instruments, with a resolution of 1 nm, XPS instrument measures and Lyman-alpha observations are combined to produce the spectra from 0.1 to 34 nm.}{The generalization of the network is tested by dividing the data sets on two groups:the training set; and, the validation set. We have found that the model error is wavelength dependent. While the model error for 24-hour forecast in the band from 115 to 180 nm is lower than 5\%, the model error can reach 20\% in the band from 180 to 310 nm. The performance of the network reduces progressively with the increase of the forecast period, which limits significantly the maximum forecast period that we can achieve with the discussed architecture.}{The model proposed allows us to predict the total and spectral solar irradiance up to three days in advance. The near real-time forecast of the total and spectral solar irradiance available at  http://www.lpc2e.cnrs-orleans.fr/$\sim$soteria .}


\authorrunning{Vieira et al.}

\titlerunning{Forecast of the Solar Irradiance}

\maketitle

%

\section{introduction}

The spectrally resolved solar irradiance (or SSI, for Solar Spectral Irradiance) is one of the key parameters for solar-terrestrial science, and in particular for space weather and space climate. The SSI in the visible and near-infrared bands is mostly absorbed in the troposphere and at ground, where it heats directly. The UV band, is predominantly absorbed in the stratosphere and above, where photolysis leads to more complex chain of mechanisms \citep{haigh07, gray10}. In particular, the Extreme-UV (EUV, 10-121 nm) is the main ionisation source of the ionosphere; any changes in the UV and EUV thus directly impact the Earth's middle and upper atmosphere. Some of their societal consequences are: increased satellite drag due to heating of the thermosphere, perturbation of ground-satellite communications due to changes in the ionospheric electron density, and on the longer-term, impact on climate change. For that reason, the continuous and long-term monitoring of the SSI has become one of the most important issues toward the quantification of the impact of solar variability on the Earth's environment \citep{lean05}.

Unfortunately, the continuous monitoring of the SSI is a major challenge as it has to be done outside of the terrestrial atmosphere, where instruments suffer from degradation and the almost total lack of in-flight calibration. First continuous observations of the SSI really started with the SORCE mission in 2003 only, as this mission, together with TIMED, was the first to provide a complete and continuous coverage of the solar spectrum, from the soft X-ray (XUV, 1-10 nm) to the near infrared. These observation have led to series of unexpected discoveries \citep{harder09}. Unfortunately SORCE now is approaching the end of of its mission life.

To compensate for this severe lack of direct observations, various empirical and semi-empirical SSI models have been developed, especially in the EUV range, which is crucial for space weather \citep{lilensten08}. All these models rely on various proxies for solar activity \citep{krivova08}. Various empirical models have been successfully using solar indices such as Mg II and f10.7  to reconstruct the SSI  \citep{lean00b, tobiska06b, lean11}. A different class of semi-empirical models relies on solar continuum images and solar magnetograms \citep{fligge98,krivova10, ball11, fontenla11}. The validity of these models has recently been challenged by the (presumably) anomalous spectral variability observed by SORCE in the near-UV \citep{haigh10}. Most of these models are based on the premise that the variability in the SSI is driven by surface magnetism only; this hypothesis has been remarkably well confirmed so far, but its validity for long-term (i.e. centennial and beyond) variations is still an open issue \citep{froehlich11, shapiro2011, vieira2011}. Note that none of them can properly reproduce short transients such as flares, for which dedicated models exist \citep{chamberlin08}. In what follows, we shall also exclude such transients and concentrate on time scales exceeding one hour.

While SSI models are crucial for understanding the physical mechanisms that are responsible for the variation of the solar radiative output \citep{domingo09}, they are also increasingly required for more applied purposes, namely for space weather applications. Unfortunately, operational requirements  bring in a number of constraints that cannot always be met by research-grade models. These include latency (the SSI must be available in near real-time) and continuity (backup options are required if there are service outages). The capacity of operational models to provide forecasts also becomes an important issue. As of today, only the commercial Solar2000 model \citep{tobiska06b} is used for operational purposes. Solar2000 uses various solar proxies to deliver SSI reconstructions covering the EUV to the near-infrared.

In the framework of the European collaborative project SOTERIA\footnote{Solar Terrestrial Interactions and Archives (2008-2011), http://soteria-space.eu/} we developed a non-commercial SSI model that meets the constraints of operational use while keeping the high physical performance of semi-empirical models that rely on solar surface magnetism. Our model, which will be detailed below, uses continuum images and magnetograms from Helioseismic and Magnetic Imager (HMI) instrument \citep{schou2011} on board of the Solar Dynamics Observatory (SDO), which are segmented into various types of solar features. The filling factors associated with these features, and their centre-to-limb position are fed into an artificial neural network that has been trained using SSI observations from SORCE instruments. The present version of the model focuses on the nowcast and short-term (1-3 days) forecast of XUV to UV ranges only (also including the TSI - Total Solar Irradiance), whereas a upcoming version will extend the range to the visible and use a flux transport model to provide one month ahead forecasts.

The paper is structured as follows. In Sect. 2 we describe the data sets employed as the models input and target. The  the neural network model, the feature extraction procedure, the preprocessing, the training, and the validation are presented in Sect. 3. In Sect. 4 we present the results. The conclusions are given in Sect. 5. 

\section{Data}
We employ solar disk magnetograms and continuum images obtained by the HMI/SDO instruments to track the evolution of bipolar magnetic regions, which are available at the HMI - AIA Joint Science Operations Center - Science Data Processing website\footnote{http://jsoc.stanford.edu/}. We use the quicklook version of the images, 1024x1024 pixels that are available with a latency of 15 minutes. Unfortunately, the calibrated images are delivered days to weeks later, and so can not be used. The images are provided in the jpg (joint photographic experts group) format in intensity levels instead of physical units. We point out that by using this format we may overestimate the area of the solar disk covered by bipolar magnetic regions because of the non-linear relationships between the magnetic field intensity of a given pixel and the actual area of that pixel covered by magnetic structures. However, the artificial neural network model is able to deal with these non-linearities. 

We have selected images with 1024x1024 pixels because of the computational limitations for the image processing and feature extraction. Although we have employed only HMI/SDO images in this analysis, observations from other instruments can be used to compute the fraction of the solar disk covered by magnetic structures. We stress that the identification of the filling factors improves significantly if images with 4096x4096 are employed. 

We employ observations of the total and spectral solar irradiance from instruments on board of the Solar Radiation and Climate Experiment (SORCE) spacecraft \citep[e.g.][]{rottman2005}. The data is available at the SORCE website\footnote{http://lasp.colorado.edu/sorce}. We use measurements from TIM instrument to monitor changes of the total solar irradiance (TSI) with a time cadence of 6 hours. We employ measurements from SIM, SOLSTICE, and XPS instruments to monitor the evolution of the irradiance in different bands of the solar spectrum, covering the XUV (0.1-10nm), part of the EUV (up to 40 nm), the FUV (121-180 nm) and the MUV (180-300 nm). The gap in the EUV will soon be filled by using data from TIMED/SEE. The time cadence of the data is 1 day. 

\section{Approach}
We employ an artificial neural network model to compute in near-real time the evolution of the solar irradiance from the distribution of magnetic active regions on the solar disk. This approach is based on the assumption that the solar irradiance variability is predominantly, if not entirely, caused by the evolution of the structure of the solar magnetic field \citep{krivova2003}. As the 3-dimensional structure of the magnetic field is imprinted on the solar surface, solar disk magnetograms and continuum images are employed to track its evolution. The emission at a given wavelength is computed from the distribution of dark and bright features on the solar disk. It is sufficient to segment the solar disk in three groups to reproduce with high precision the evolution of the total and spectral solar irradiance. The groups needed for capturing the salient features of the solar spectral variability are: the quiet sun, sunspots (umbra and penumbra), and bright magnetic structures (faculae and the network). 

Models such as the SATIRE (Spectral And Total Irradiance Reconstruction) models are also based on the same assumption and have been successfully employed to compute the evolution of the solar irradiance from days to millennia \cite{krivova2003}. However, while SATIRE models employ the intensities of each atmospheric component to compute the emission at a given wavelength, we use a more empirical and data-driven approach based on an artificial neural network (ANN). The main advantage of using an ANN instead of the intensities of each atmospheric component is the flexibility to recognize and predict temporal patterns from near-real time solar disk magnetograms and intensity images. The model, however, needs to be trained with real data so our assumption here is that the spectra from SORCE are indeed the true representation of the SSI.

\subsection{Description of the Artificiall Neural network mode and Datal}
\begin{figure}
 \centerline{\includegraphics[width=1.0\textwidth,clip=true, viewport=0.2cm 0.4cm 17cm 10.0cm]{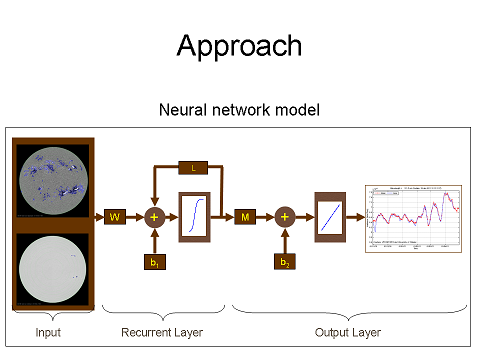}}
\caption{Schematic representation of the artificial neural network architecture. See text for a detailed explanation of the network structured.}
\label{Fig_ann}
\end{figure}

We employ a Layer-Recurrent Network (LRN) to model the complex relationships between the evolution of the bipolar magnetic structures (input) and the solar irradiance (output). We follow the original structure proposed by \cite{elman90}. Figure \ref{Fig_ann} presents a schematic representation of the LRN architecture. The network has two layers with one neuron in each layer. While the network uses a nonlinear transfer function (hyperbolic tangent - $tanh$) for the first layer, a linear transfer function is employed for the output layer. Additionally, there is a feedback loop, with a single delay, around the first layer. This feedback introduces memory in the system and helps the stabilizing the spectra when the inputs suffer from outliers. 

The output of the first layer ($a_{1,i}$) at the discrete  time instant $t_i$ is computed taking into account weighted input (${\boldsymbol W} {\boldsymbol p}$), the bias ($b_1$) and the weighted feedback ($L a_{1,i-1}$). The input of the network is the n-element vector $p$. The elements of  the input vector $p_j$ are multiplied by weights $w_j$. The weighted values are then summed.  In this way, we can express mathematically the output of the first layer as   
\begin{equation}
a_{1,i} = tanh({\boldsymbol W} {\boldsymbol p} + {\boldsymbol L} a_{1,i-1} + b_1)
\end{equation}
The output of the second layer ($a_{1,i}$), which is the irradiance at a given wavelength and at a given discrete time ($I(\lambda,t_i)$), is represented by a linear combination
\begin{equation}
a_{2,i} ={\boldsymbol M} a_{1,i} + b_2
\end{equation}

In principle, solar disk magnetograms and intensity images can be employed directly as the input of the network. However, this would imply in a very large number of coefficients to be determined, which is not computationally efficient. One alternative is the transformation of the input images into a set of features. This process, which is known as feature extraction, is a form of dimensionality reduction. In our case, the simplest way to proceed is the determination of the fraction of the disk covered by magnetic structures, the filling factors. The algorithm employed to determine the filling factors is described in detail in the next section. 

The coefficients $\boldsymbol W$, $\boldsymbol L$, $\boldsymbol M$, $b_1$, and $b_2$ are determined by minimizing a combination of squared errors and weights \cite{MacKay1992}. This process is also known as Bayesian regularization. In this way, we determine the set of coefficients to produce a network that generalizes properly the relations between the input and the output data. 

The algorithm described above can be applied for nowcast as well as for short-term forecast up to two days. Here, we define nowcast as a short-term forecast out to six hours. 

\subsection{Feature extraction procedure}

\begin{figure*}
\centering
\includegraphics[width=1.0\textwidth,clip=true, viewport=1.8cm 2.5cm 30cm 30cm]{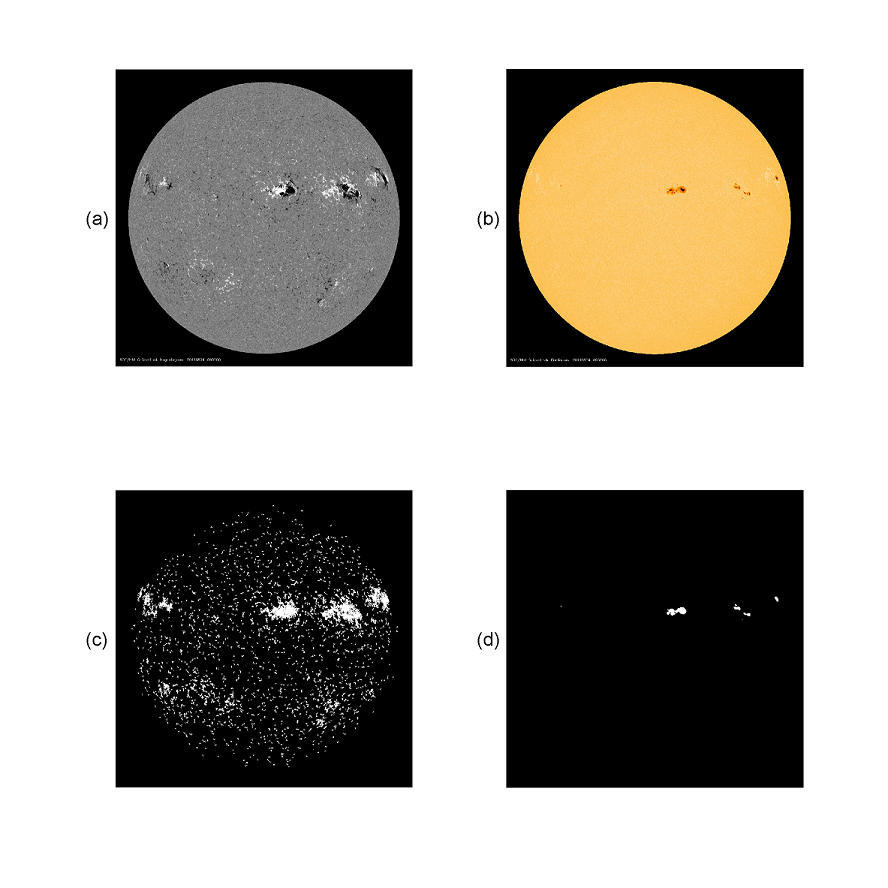}
\caption{ SDO/HMI solar disk line of sight magnetogram (a) continuum intensity image (b) measured on 04-Aug-2011 09:00:00. (c) Bipolar regions identified in the magnetogram. (d) Sunspots identified in the continuum image.}
\label{FigHMI1}
\end{figure*}

Here we employ solar disk magnetograms and continuum images to identify the magnetic active regions and sunspots, respectively. An example of such observations is provided in Figures \ref{FigHMI1}a-b, which show the observations of the solar disk obtained by HMI instruments on 04-Aug-2011 at 9:00 UT. Large magnetic active regions and sunspots are present in the northern hemisphere. The feature extraction procedure consists of the identification of the quiet Sun, magnetic active regions, and sunspots. 

The method to identify magnetic active regions in the disk magnetograms includes the following steps: (a) identification of the disk pixels; (b) segmentation of the image; (c) connected-component labelling; (d) computation of the area of each object identified in the binary image; (e) removal of small area objects. 

The identification of the disk pixels is made taking into account that most of the pixels outside of the solar disk correspond to black pixels ($I_m[i,j] = 0$). In this way, we segment the image employing a simple threshold. Note that by applying this threshold the annotation's pixels outside of the disk will also be identified as disk pixels. In order to deal with this problem, we search for the feature with the largest area.

To distinguish the magnetic active regions from the quiet Sun, we segment the disk magnetograms ($X_{m,p}$) employing by a threshold

\begin{equation}
X_{m,s}[i,j] = 
\begin{cases}
0 & \text{if $|X_{m,p}[i,j] - X_{m,0}| < X_{m,th}$,} \\
1 & \text{if $|X_{m,p}[i,j] - X_{m,0}| \ge X_{m,th}$.} 
\end{cases}
\end{equation}
where $X_{m,0}$ is the intensity level that corresponds to 0 Gauss and $X_{m,th}$ is the threshold. One example of the binary image resulting from the segmentation is shown in Figure \ref{FigHMI1}c. White regions represent magnetic active regions, the quiet Sun is shown in black. 
Next, we perform a connected-component labeling, which is a method for identifying each object in a binary image. The connectivity is four (4), which means that we search for 4-connected neighborhood. The area of each object, which is employed to classify the objects, is then computed. We remove objects with small areas (less than 10 pixels).

We identify sunspots (umbrae and penumbrae) in the solar disk intensity images. The procedure is similar to the one used to identify magnetic active regions. However, as sunspots are dark features in the solar disk, we search for pixels that are below a given threshold. In order to distinguish umbrae and penumbrae, we apply two thresholds. 
\begin{equation}
X_{c,s}[i,j] = 
\begin{cases}
0 & \text{if $(X_{c,p}[i,j] - X_{c,0}) > X_{c,th_p}$,} \\
1 & \text{if $(X_{c,p}[i,j] - X_{c,0}) \le X_{c,th_p}$ and $(X_{c,p}[i,j] - X_{c,0}) > X_{c,th_u}$,} \\
2 & \text{if $(X_{c,p}[i,j] - X_{c,0}) \le X_{c,th_u}$.} 
\end{cases}
\end{equation}
where $X_{c,0}$ is the reference level and $X_{c,th_u}$ and $X_{c,th_u}$ are the thresholds for umbrae and penumbrae, respectively. An example of a binary image obtained is presented in \ref{FigHMI1}d. The sunspots identified are presented in white. 

In Figure \ref{FigHMI1}c we observe that bipolar regions occur in a continuum size spectrum \cite{harvey1993}. However, it is convinient to divide the spectrum into active regions and ephemeral regions. Here we divide the spectra into four classes according to the filling factors of individual structures, i.e., the fraction of the solar disc covered by an individual magnetic structure. The classes are determined from the empirical cumulative distribution function (ECDF) of the features identified from Sep/2010 to Dec/2010. Note that the ithe thresholds  employed for the segmentation of the imagess are fixed taking into account that the noise should be removed. The threshold values affect the distribution of the filing factors of individual structures. 

\begin{figure*}
\centering
\includegraphics[width=1.0\textwidth,clip=true, viewport=1.0cm 0.0cm 19cm 10cm]{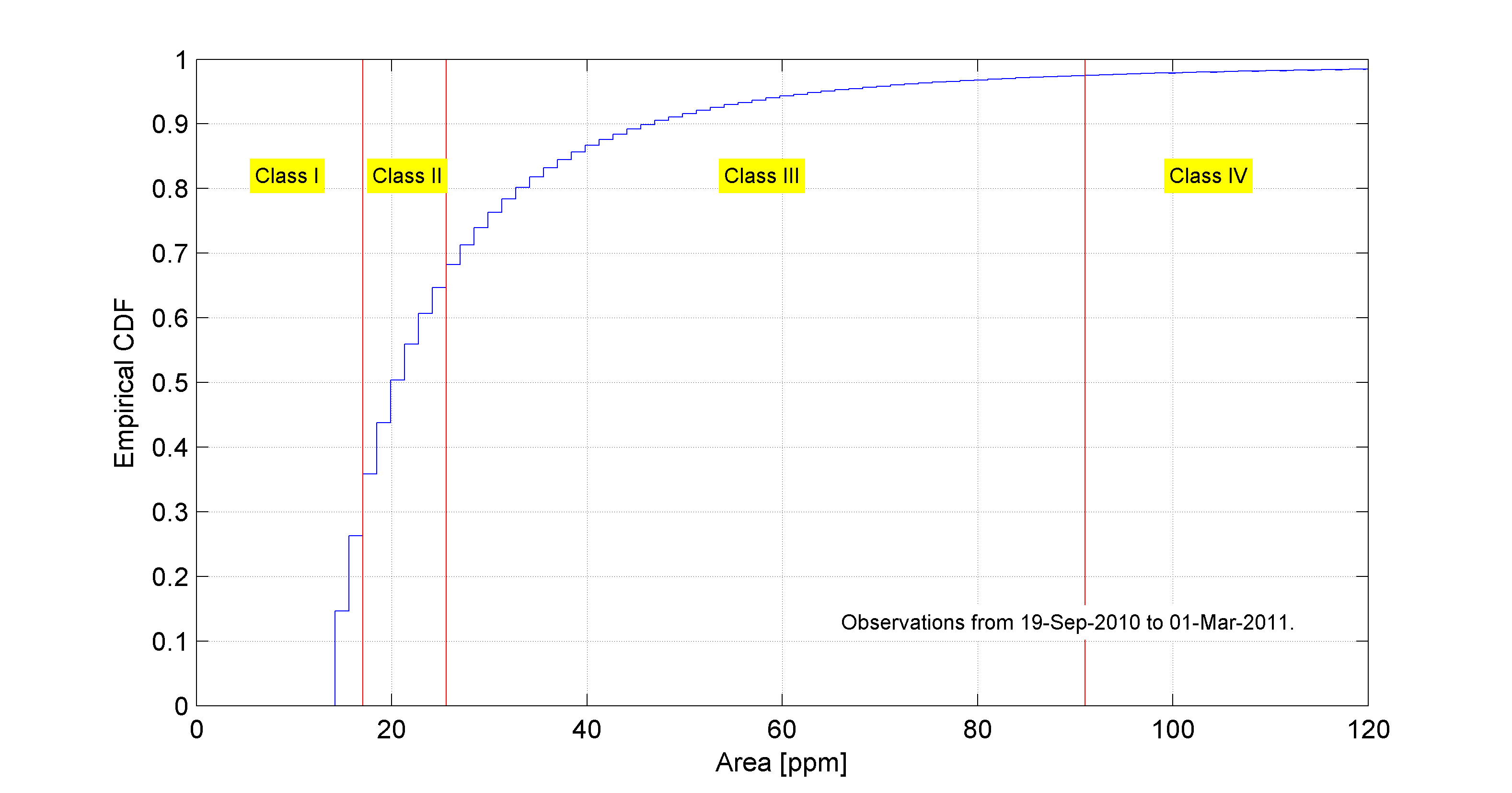}
\caption{ Empirical cumulative distribution function of the filling filing factors of bipolar magnetic active regions identified in the magnetograms from 19-Set-2010 to 01-Mar-2010. The red lines indicate the filling factors at 0.33, 0.66, and 0.97. The four classes are marked in the figure.}
\label{Fig_ecdf}
\end{figure*}

\begin{figure*}
\centering
\includegraphics[width=0.8\textwidth,clip=true, viewport=0.0cm 1.0cm 30cm 35cm]{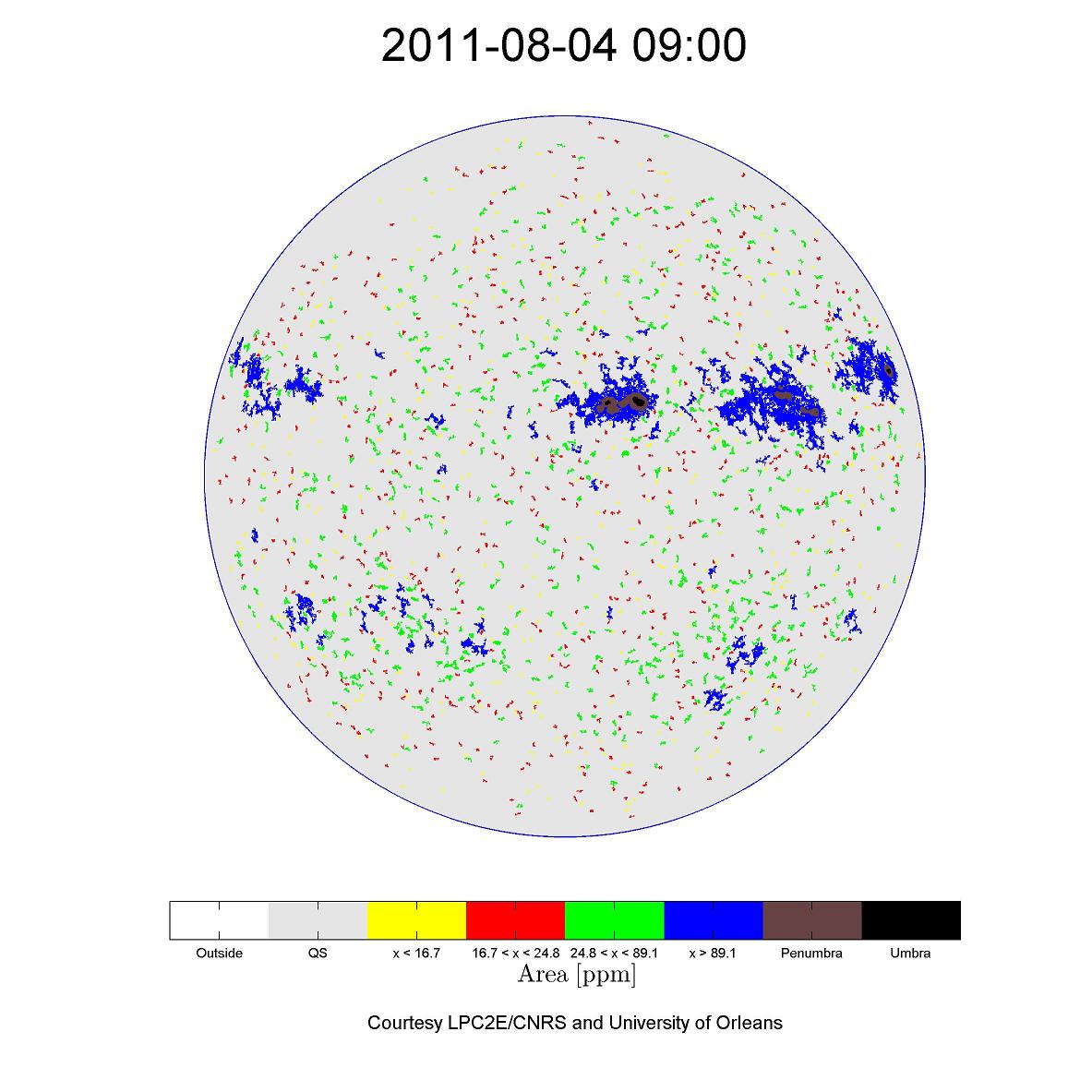}
\caption{ Distribution of bipolar magnetic regions identified on 04-Aug-2011 09:00:00. The color scheme indicates the four classes of active regions considered in this work. The umbrae and penumbrae regions are also indicated in the figure. The quite sun is shown in gray.}
\label{Mask}
\end{figure*}

In Figure \ref{FigHMI1}c we observe that bipolar regions occur in a continuum size spectrum \cite{harvey1993}. However, it is convenient to consider separately the spectral contribution from active regions and ephemeral regions. Here we divide the spectra into four classes according to the filling factors of individual structures, i.e. the fraction of the solar disc covered by an individual magnetic structure. These classes are determined from the empirical cumulative distribution function (ECDF) of the features identified from Sep/2010 to Dec/2010. Figure \ref{Fig_ecdf} presents the ECDF obtained (blue line). The boundaries between the four classes are defined approximately at the probability levels: 33.3\%, 66.6\%, and 97\%. Table \ref{table1} shows the resulting classification according to the filling factors of the structures. Following this classification scheme, we produce an image mask in which the active regions and the sunspots are represented. Figure \ref{Mask} displays an example of the image mask produce from the magnetogram and intensity image of 04-Aug-2011 at 09:00:00. 

\begin{table}[h!b]
\caption{Classification scheme of the bipolar magnetic structures according to the filling factors.}
\begin{tabular}{c c }
Class & Filling Factors (ppm) \\
\hline
 I    &  $\alpha \leq 16.7$  \\
 II  & $ 16.7 < \alpha \leq 24.8$  \\
 III & $ 24.8 < \alpha \leq 89.1$  \\
 IV  & $\alpha > 89.1$  \\
\hline
\end{tabular}
\label{table1}
\end{table}

\begin{figure}
\centering
\includegraphics[width=0.5\textwidth,clip=true, viewport=2.9cm 20cm 10cm 28cm]{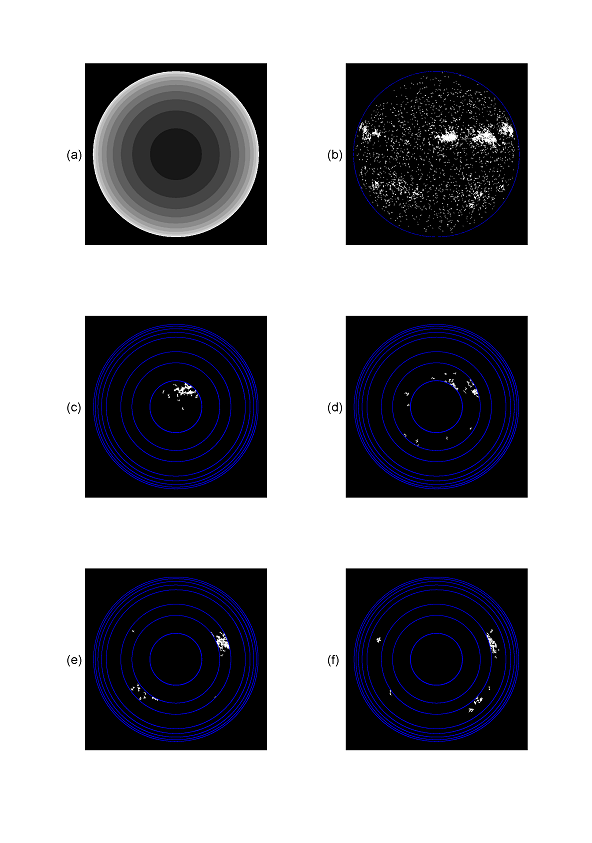}
\caption{Distribution of the concentric rings employed to take into the account the center-to-limb contrast of the bipolar regions.}
\label{Fig_mu_rings}
\end{figure}

The contribution of the solar features to the solar irradiance also depends on the position of the features on the solar disk. In order to take this in to account, we compute the area  covered by bipolar features in concentric rings. These rings are determined according to the heliographic angle ($\mu$). Figure \ref{Fig_mu_rings} shows the eleven (11) rings considered in this work. As we will show later, there is no need for increasing that number.

The input vector ($p$) of the network is defined as the filling factors of the 10 inner rings of each class considered. Following a common pratice, we normalize the input time series proportionally to the standard deviation before they enter in the neural network. Note that the thresholds employed for the segmentation of the images are fixed taking into account that the noise should be removed. Although the threshold values affect the distribution of the filing factors of individual structures, the training procedure accommodate the ANN coefficients in order to generalize properly the output. 

\section{Results and Discussions}

\subsection{Evolution of the Filling factors}

\begin{figure*}
\centering
\includegraphics[width=1.0\textwidth,clip=true, viewport=0.2cm 4.0cm 19cm 20cm]{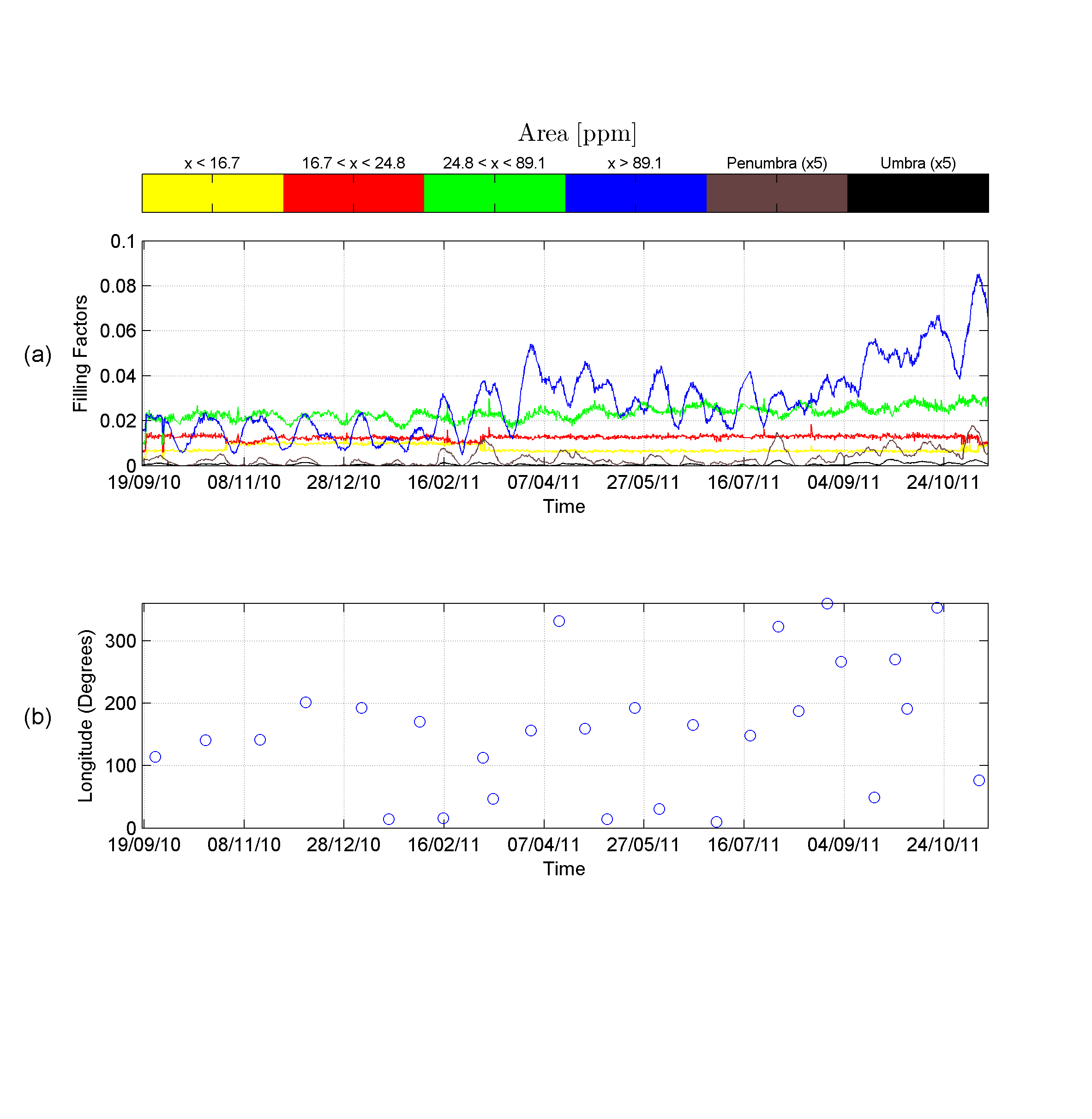}
\caption{(a) Evolution of the filling factor of the several components. The evolution of large active regions is presented in blue, while the evolution of regions with filling factors between 24.8 and 89.1 ppm are shown in green. The red and yellow lines display the evolution of regions with filling factors lower than 24.8 ppm. The evolution of the penumbrae and umbrae are presented in brown and black, respectively. (b) Heliographycal longitude of the disk center during the peaks of filling factors of large magnetic active regions.}
\label{fig_small_areas}
\end{figure*}

Figure \ref{fig_small_areas}a shows the evolution of the various filling factors from Sep/2010 to Oct/2011. Each line presents the fraction of the solar disk covered by structures that belong to one class, i.e. the filling factors of each class. The yellow, red, and green lines display the evolution of ephemeral regions (ER) that are members of the Classes I, II, and III, respectively. The blue line exhibits the evolution of active regions (AR), which are structures with filling factors larger than 89.1 ppm (Class IV). The brown and black lines show the evolution of the sunspots penumbrae and umbrae, respectively. Note that for a better visualization, the filling factors of penumbrae and umbrae were multiplied by a factor of five (5). 

We find that the filling factors of Classes I and II structures do not present trends during the period considered. However, an increase of the filling factors of Class I structures in relation to the average value is observed from 29-Oct-2010 to 07-Mar-2011. Changes of the filling factors of Class II structures are observed near the boundaries of this interval. These changes appear as decreases of the filling factors of Class II structures from 29-Oct-2010 to 10-Nov-2010 and from 15-Feb-2011 to 07-Mar-2011. We speculate that these variations are due to small changes in the calibration of the magnetograms or the mapping of the quicklook images. In this interval, we did not observe any such discontinuity in the filling factors of Classes III and IV structures. We also find that oscillations with periods longer than few days are not present in the time series of Classes I and II, which suggests that they are quasi-uniformly distributed on the solar surface. Although this result has to be confirmed with higher resolution and calibrated magnetograms and intensity images, it implies that the evolution of small size structures ($\alpha < 24.8$ ppm) does not contribute significantly to medium-term changes of the solar surface magnetic flux (i.e. on time scales of months). Consequently, long-term changes of the solar surface magnetic flux seem to be due only to the evolution of structures with filling factors larger than 24.8 ppm. Furthermore, assuming that changes of the solar irradiance are due only to the evolution of the magnetic field structure, the long-term evolution of the solar irradiance may then be constrained only by the evolution of Class III and IV structures. 

A quasi-periodic 27-day modulation of the filling factors of active regions is observed from the middle of Set/2010 to the middle of Jan/2011. The maxima and minima values remained at the same level during this period. The heliographic longitude of the disk center during the peaks of the filling factors of active regions is in the sector between 100 and 200 degrees (see Fig. \ref{fig_small_areas}b). After a short period of low activity, the area covered by active regions increased approximately four (4) times from Jan/2011 to Nov/2011. The increase of the complexity of the signal after Jan/2011 reflects the emergence of active regions in the longitudinal sector from -50 to 50 degrees. In the same way, the filling factors of the Class III structures increased by about 50\% from Sep/2010 to Oct/2011. These patterns for the increase of the area covered by active and ephemeral regions are in agreement with previous studies suggesting that ephemeral regions evolve in cycles that are longer than sunspot cycles. Consequently, as the ephemeral regions evolve in cycles that overlap significantly, the variation through the 11-year cycles is lower than the variation of the area covered by active regions. 
 
\begin{figure*}
\centering
\includegraphics[width=1.0\textwidth,clip=true, viewport=0.0cm 1.5cm 20cm 29cm]{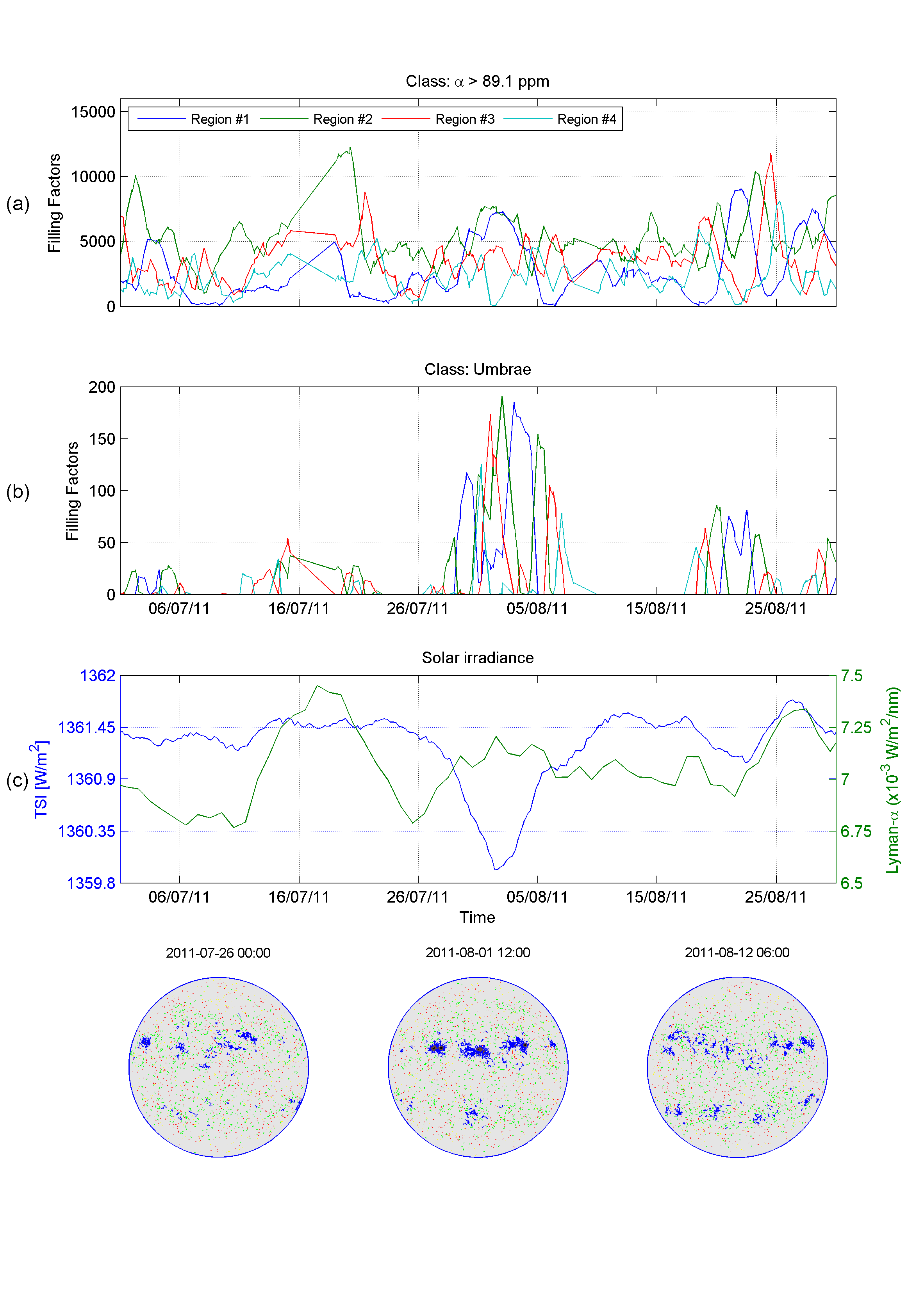}
\caption{Evolution of the filling factors of the four inner rings. Panels (a) and (b) present the evolution of the structures with filling factors lower than 16.7 ppm and between 16.7 and 24.8 ppm, respectively. The evolution of large structures is presented in panels (c) and (d), while the evolution of penumbrae and umbrae is presented in panels (e) and (f).}
\label{Fig_alpha_mu}
\end{figure*}

In addition to the presence of active regions on the solar disk, its spatial distribution  affects directly the variability of the solar irradiance. Figure \ref{Fig_alpha_mu} shows the evolution of the filling factors of the four (4) inner rings. Panels (a) to (d) present the evolution of the filling factor of Classes I, II, III, and IV structures, respectively. The evolution of the sunspots penumbrae and umbrae are displayed in panels (e) and (f), respectively. While the changes of the area covered by large active regions and sunspot groups are easily noticeable when considering individual rings (see Fig. \ref{Fig_alpha_mu}d-f), the filling factors of Classes I and II structures do not present any significant modulation. The filling factors of Class III structures are also affected by the presence of active regions, but these structures are observed in a wider region of the solar disk. 

\begin{figure*}
\centering
\includegraphics[width=1.0\textwidth,clip=true, viewport=0.0cm 1.5cm 19cm 29cm]{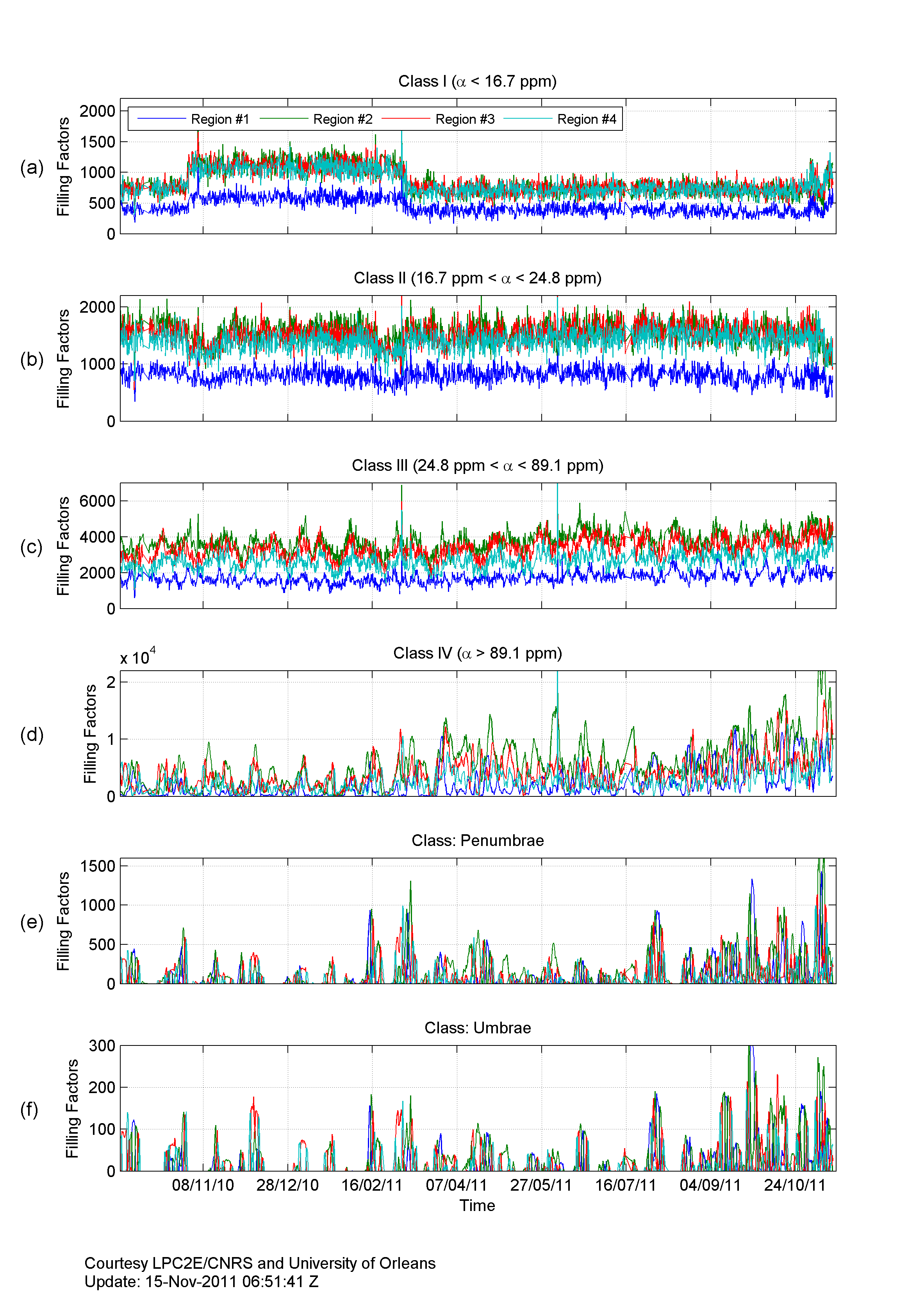}
\caption{Evolution of the filling factors of the four inner rings. Panels (a) and (b) present the evolution of the structures with filling factors lower than 16.7 ppm and between 16.7 and 24.8 ppm, respectively. The evolution of large structures is presented in panels (c) and (d), while the evolution of penumbrae and umbrae is presented in panels (e) and (f).}
\label{Fig_alpha_mu_zoom}
\end{figure*}

Figure \ref{Fig_alpha_mu_zoom} exhibits a comparison between the evolution of active regions and the solar irradiance. Panels (a) and (b) present the filling factors for large active regions and sunspot umbrae, respectively. Panel (c) displays the evolution of the total solar irradiance (blue line) and Lyman-$\alpha$ emission. The distribution of active regions on the solar disk prior, during, and after the passage of the sunspot groups are presented in the bottom of the figure. Note that during the transit of sunspots on the solar disk the total solar irradiance reduces significantly as expected. The transit of the active regions from their appearance on 2011-07-26 to the maximum on 2011-08-01 is observed as a progressive increase of the filling factors from the limb to the center rings. At the same time, the TSI decreases progressively, reaching its minimum value around 2011-08-01 12:00 when three (3) large active regions in the northern hemisphere are near the disk center. A gradual decrease of the area covered by the active regions/sunspots and the concomitant increase of the TSI occurs from the 2011-08-01 to 2011-08-10. However, as the pattern of the Lyman-$\alpha$ emission illustrates, the presence of active regions produce different patterns in different bands of the spectrum. In contrast to the decrease of the TSI during the passage of the active regions, a clear enhancement of the Lyman-$\alpha$ emission is observed from 2011-07-26 to 2011-08-10. A large enhancement of the Lyman-$\alpha$ emission is also observed from 2011-07-11 to 2011-07-26 without a concomitant decrease of the TSI. In this way, the feature extraction procedure is able to identify the presence and the distribution of the active regions/sunspots that are needed to compute the evolution of the solar irradiance employing a neural network.  

\subsection{Short-term forecast of solar irradiance}

As discussed in Sect. 3, the coefficients of the network are obtained by comparing the output of the model and observations of the solar irradiance made by instruments onboard of SORCE spacecraft. Here we focus on two regions of the spectra that are covered by SORCE instruments. While the range from 115 nm to 310 nm is covered by the two SOLSTICE instruments, with a resolution of 1 nm, the XPS instrument measures spectra from 0.1 to 34 nm. 

\begin{figure*}
\centering
\includegraphics[width=1.0\textwidth,clip=true, viewport=0.0cm 2.00cm 19cm 28cm]{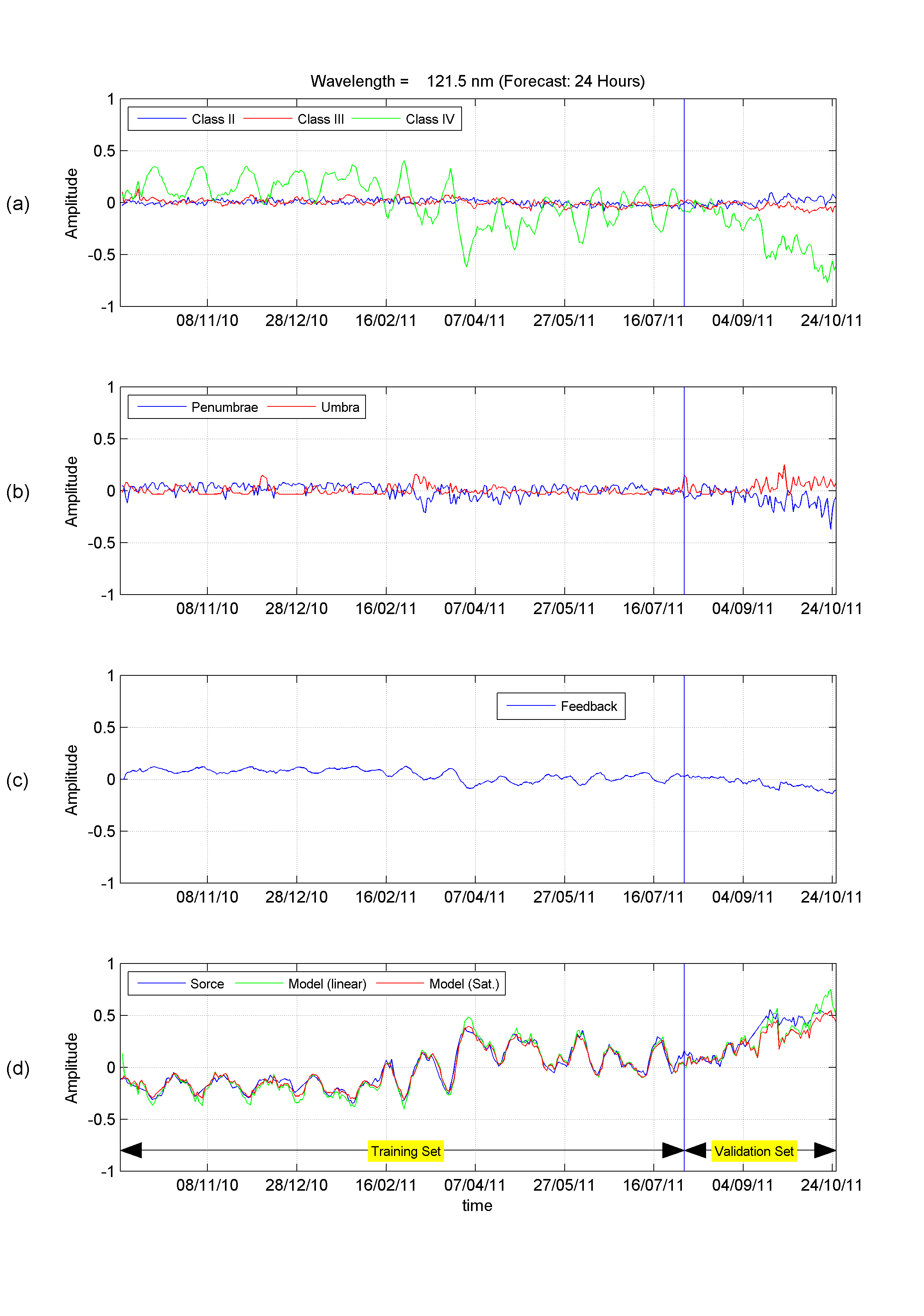}
\caption{Example of a 24-hour forecast training section of the neural network model for the wavelength band centered at 121.5 nm (Lyman alpha). The upper panel shows the evolution of the contributions from classes II, III, and IV. The Panel (b) presents the evolution of umbrae and penumbrae. The panel (c) shows the evolution of the feedback of the first layer to the input of this layer. The panel (d) shows a comparison between the output of the model (red curve) and the observations. For reference, the output of the same model, but with a linear transfer function in the hidden layer is presented. The training and validation sets are marked in the panel (d).}
\label{figure_fuv_007_contrib}
\end{figure*}

Figure \ref{figure_fuv_007_contrib} shows a comparison between the observations of the irradiance at 121.5 nm measured by the SOLSTICE instrument and the output of the 24-hour forecast model. The red and blue lines in the upper panel present the contribution of ephemeral regions, Classes II and III, respectively. The green line displays the contribution of active regions (Class IV). Figure \ref{figure_fuv_007_contrib}b displays the contribution of umbrae (red line) and penumbrae (blue line). The contributions are computed by multiplying each element of the input vector by the weights $w_i$. The weighted elements of each class are them summed and the total contribution of each class is obtained. Figure \ref{figure_fuv_007_contrib}c shows the feedback contribution (${ \boldsymbol L} a_{1,i-1}$). The large this value is (in absolute term) the more the ANN relies on past values to estimate the present one. That is, a large feedback implies a reconstruction based on persistence. In this particular example (Fig. \ref{figure_fuv_007_contrib}c), the feedback is negligible. Figure \ref{figure_fuv_007_contrib}d presents the time series from SOLSTICE (blue line), the neural network output (red line), and the neural network output with a linear transfer function for the first layer (green line). The training (80\%) and validation (20\%) sets are indicated in the figure. The model reproduces adequately the variability of the training set as well as the validation set, which indicates that the model properly generalizes the relations between the distribution of bipolar magnetic features on the solar disk and the Lyman-$\alpha$ emission. Note also that the output of the linear model, in which each neuron has a linear response, in most cases, performs as well as the nonlinear one, except with larger excursions of outliers. 

\begin{figure*}
\centering
\includegraphics[width=1.0\textwidth,clip=true, viewport=1.0cm 0.0cm 19cm 12cm]{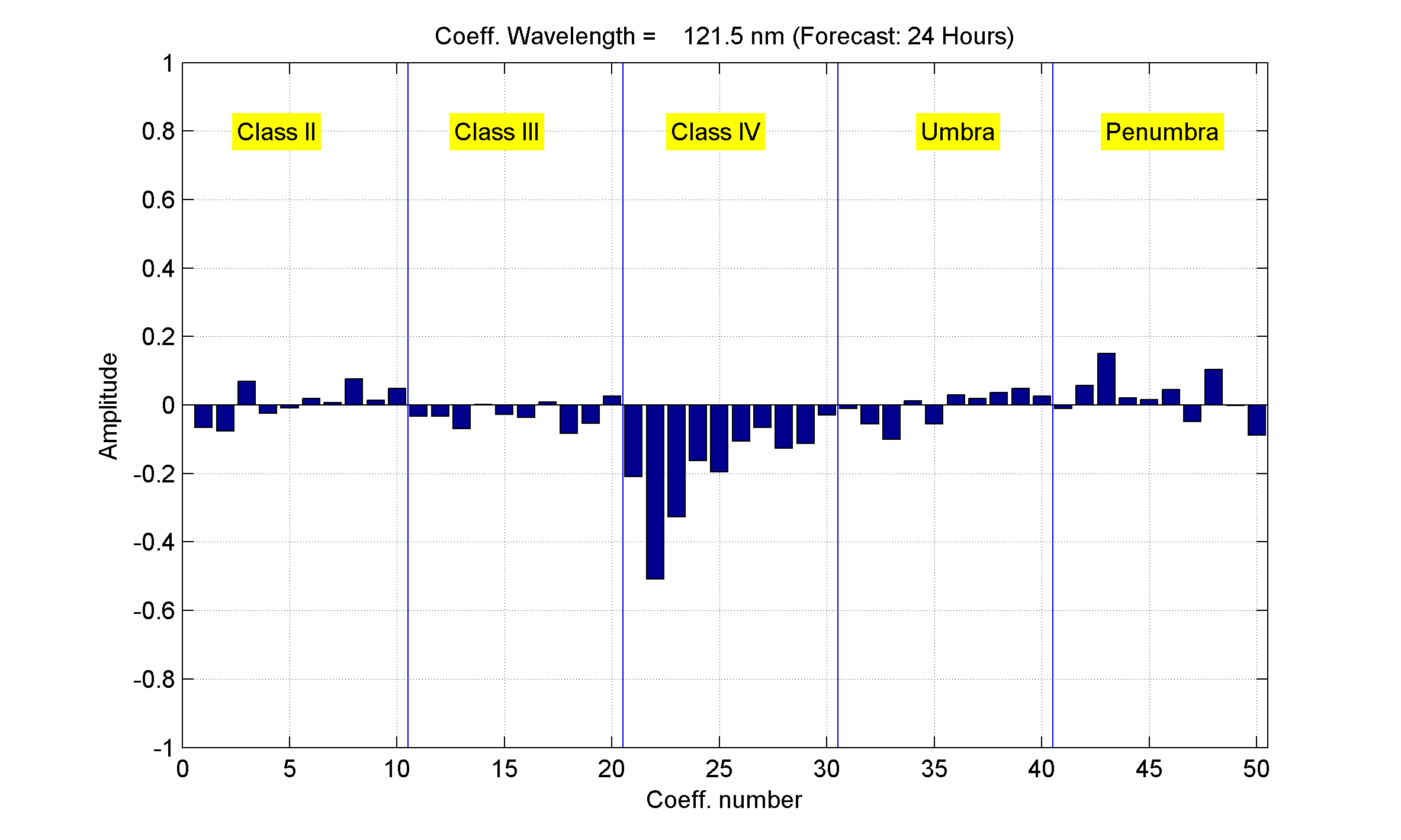}
\caption{Coefficients $W$ for the wavelength of 121.5 nm (Lyman alpha). The coefficients for each class are indicated in the figure. The evolution of the inner rings is on the left for each class.}
\label{Fig_fuv_007_coeff}
\end{figure*}

The coefficients of the model are shown in Figure \ref{Fig_fuv_007_coeff}. The coefficients of each class considered are indicated as well as the coefficient corresponding to the inner ring. In this example, most of the contribution for the evolution of the irradiance is due to the evolution of the large active regions. As expected, the major contribution to the variability of the Lyman-$\alpha$ emission is due to the evolution of active regions (Class IV), although the feedback also contributes. Not surprisingly, this contribution mainly comes from active regions that are near the center of the disk. This property is, of course, wavelength dependent. Incidentally, because the model is data driven, we now can use it to infer properties about the radial contribution of specific features for each wavelength. This opens interesting perspectives that will be investigated in a forthcoming publication. 

\begin{figure*}
\centering
\includegraphics[width=1.0\textwidth,clip=true, viewport=0.0cm 1.0cm 20cm 7.5cm]{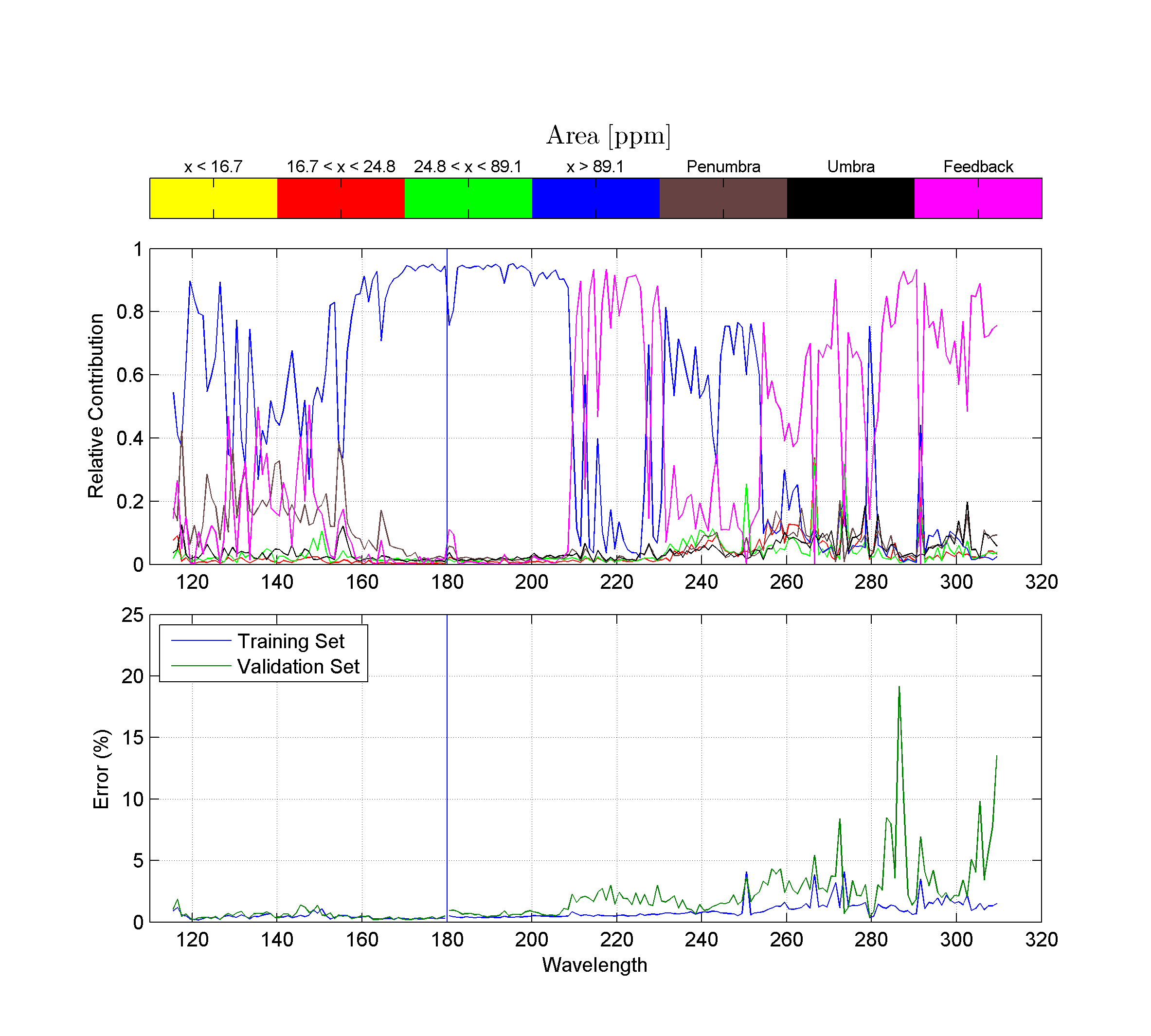}
\caption{Percentual error of 24-hour forecast model from 115 to 310 nm. The blue line presents the model error for the training set while the green line shows the model error for the validation set.}
\label{figure_mse}
\end{figure*}

The percentual difference between the output of the for 24-hour forecast model and the observations, the model error, is presented in Fig. \ref{figure_mse} for the training (blue line) and validation (green line) sets from 115 to 310 nm. It is noticeable that the model error of the MUV region of the spectra is higher than in the FUV.  The model error for the XPS region of the spectra, which is not shown in the figure, is comparable to the error of the FUV region. 

\begin{figure*}
\centering
\includegraphics[width=1.0\textwidth,clip=true, viewport=0.0cm 2.00cm 15cm 22cm]{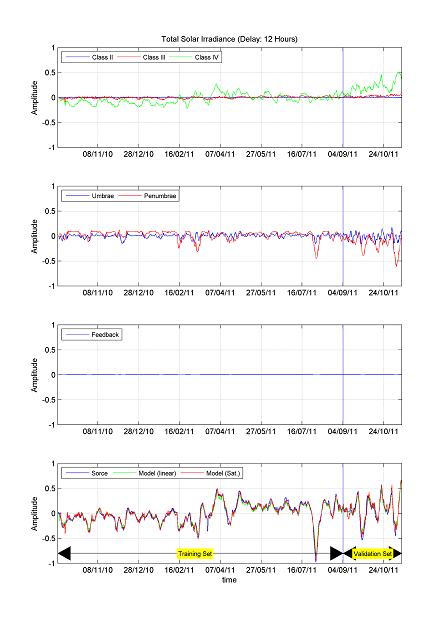}
\caption{ Example of a 12-hour forecast training section of the neural network model for the total solar irradiance. The structure of the figure is the same of the Fig. \ref{figure_fuv_007_contrib}. }
\label{FigTSI_12}
\end{figure*}

\begin{figure*}
\centering
\includegraphics[width=1.0\textwidth,clip=true, viewport=0.0cm 2.0cm 20cm 29cm]{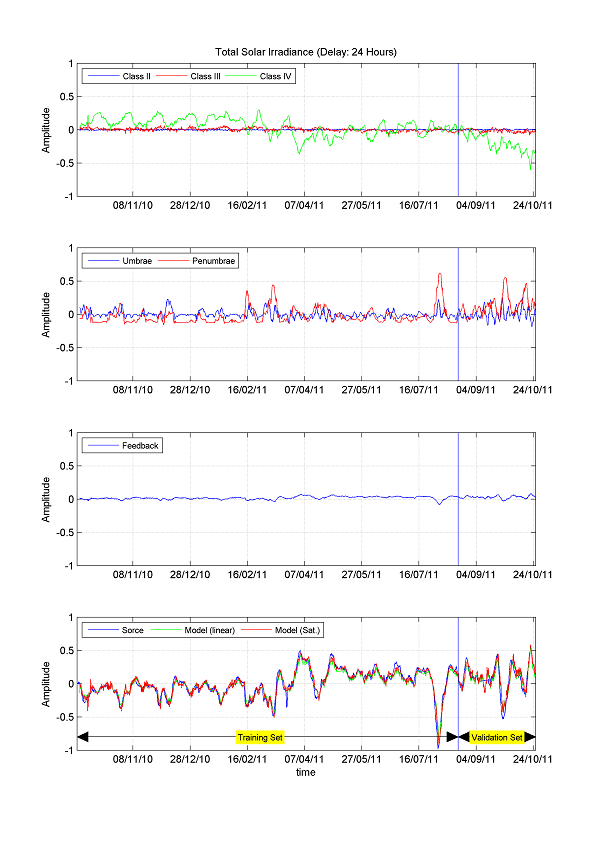}
\caption{Example of a 24-hour forecast training section of the neural network model for the total solar irradiance. The structure of the figure is the same of the Fig. \ref{figure_fuv_007_contrib}.}
\label{FigTSI_24}
\end{figure*}

\begin{figure*}
\centering
\includegraphics[width=1.0\textwidth,clip=true, viewport=0.0cm 2.0cm 20cm 29cm]{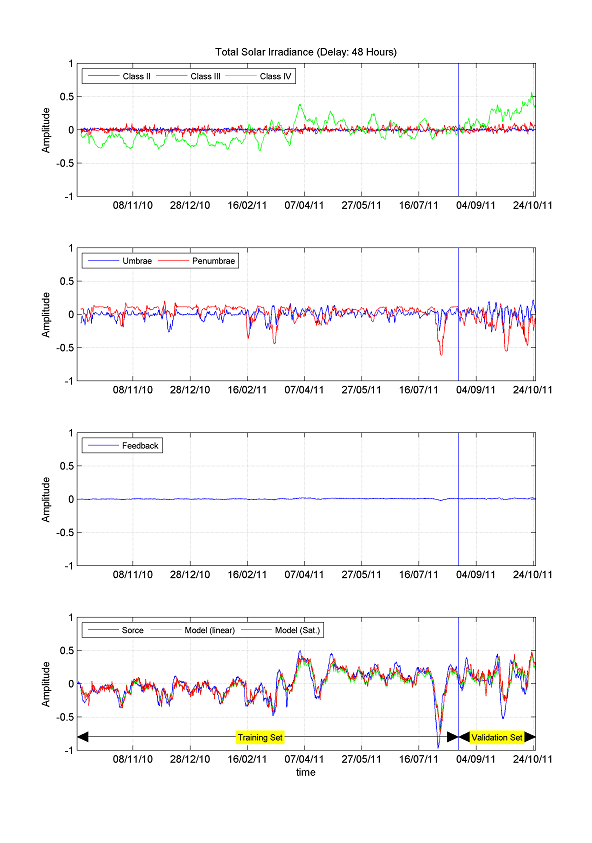}
\caption{ Example of a 48-hour forecast training section of the neural network model for the total solar irradiance. The structure of the figure is the same of the Fig. \ref{figure_fuv_007_contrib}.}
\label{FigTSI_48}
\end{figure*}

\begin{figure*}
\centering
\includegraphics[width=1.0\textwidth,clip=true, viewport=0.0cm 2.0cm 20cm 29cm]{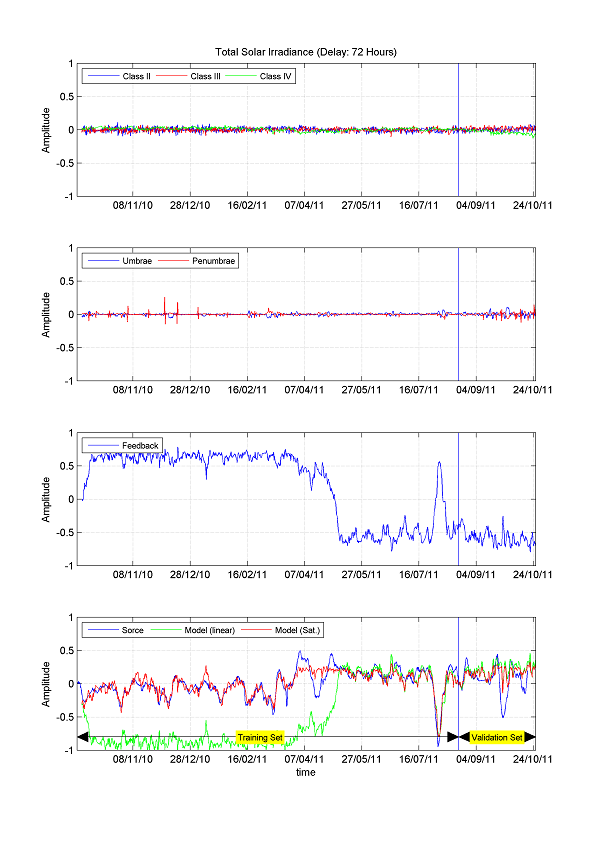}
\caption{Example of a 72-hour forecast training section of the neural network model for the total solar irradiance. The structure of the figure is the same of the Fig. \ref{figure_fuv_007_contrib}.}
\label{FigTSI_72}
\end{figure*}

Examples of training sections for the total solar irradiance for forecast periods from 12 hours to 72 hours are displayed in Figures \ref{FigTSI_12}-\ref{FigTSI_72}. The structure of these figures is the same of the Fig. \ref{figure_fuv_007_contrib}. As expected, the model error increases as the forecast period, i.e. the time interval for which a forecast is made, increases. Moreover, while the contribution of the feedback for predictions up to 48 hours is negligible, the contribution of the feedback for 72-hour forecast is significant. This indicates that, as predictable, 72-hour forecasts rely heavily on past observations. 

The increase of the model error in function of the forecast period and the progressive dependence on past observations are related to the emergence and decay of the magnetic active regions as well as their transit on the solar disk as seen from our vantage point of view. The evolution of the magnetic active regions affects the structure of the solar atmosphere and its electromagnetic emission, which are not predictable employing the ANN architecture discussed in this paper. Additionally, the solar surface does not rotate as a solid body, which is an implicit assumption in this architecture. These sources of errors can be reduced employing a surface magnetic flux transport model as briefly described in the next subsection.  

\subsection{Subsequent steps toward a medium-term  forecast}

In addition to short-term forecasts, predictions of the solar irradiance on time scales of months are needed for space weather applications, such as the evaluation of the upper atmosphere and thermosphere conditions.  

In principle, the artificial neural network described in Sect. 3 can be adapted for the forecast of the solar activity on the time scales of months. The adaptation would involve the increase of the complexity of the model by increasing the number of neurons and/or layers. One alternative is predicting the distribution of the magnetic structures by employing a solar surface magnetic flux transport model \citep{jiang2010} and them apply the algorithm described in Sect. 3. 

In a magnetic flux transport model, the variation of the radial component of the surface magnetic flux can by computed by

\begin{eqnarray*}
 \frac{\partial B_r}{\partial t} & = & + \sum S(\theta,\phi, t, Class_i) -\omega(\theta) \frac{\partial B_r}{\partial \phi} - \frac{1}{R_{sun} \sin(\theta)}\frac{\partial [v(\theta)B_r \sin(\theta)]}{\partial \theta} + \\
&& \frac{\eta_{h}}{R_{sun}^2} \left [ \frac{1}{\sin(\theta)} \frac{\partial}{\partial \theta}\left ( \sin(\theta) \frac{\partial B_r}{\partial \theta}   \right ) + \frac{1}{\sin^2(\theta)}\frac{\partial^2 B_r}{\partial \phi^2}  \right ]  - D_r(\eta_r B_r) \\
\end{eqnarray*}
where $S(\theta,\phi, Class_i)$ is the source term of the different classes and $D_r(\eta_r B_r)$ is the decay term parameterizing the radial diffusion of the magnetic field. $\eta_h$ and $\eta_r$ are the horizontal and radial diffusivity, respectively. While $v(\theta)$ is the meridional flow, $\omega(\theta)$ is the latitudinal differential rotation. 

\begin{figure*}
\centering
\includegraphics[width=1.0\textwidth,clip=true, viewport=0cm 11.0cm 18cm 22cm]{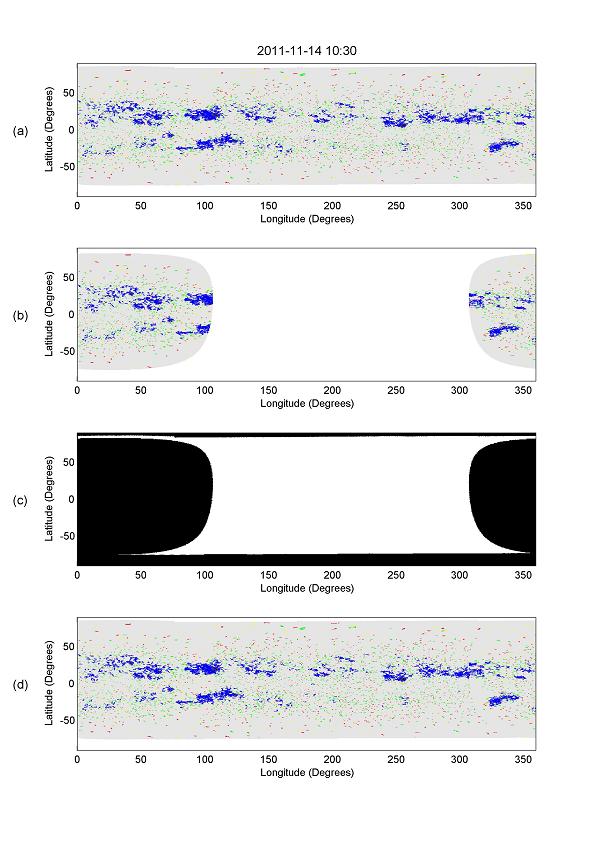}
\caption{Charts of the distribution of the surface magnetic field computed from the disk magnetograms measured during one solar rotation. The visible disk on 2011-11-14 at 10:30 is presented on panel (b). The color scheme follows the one for Fig. \ref{Mask}}
\label{FigMask_projected}
\end{figure*}

The starting point are synoptic-like charts of the distribution of the surface magnetic field computed from the disk magnetograms measured during one solar rotation. Figure \ref{FigMask_projected}b shows an example of the distribution of magnetic structures for one solar rotation. The visible disk on 2011-11-27 at 33:30 is presented on panel (b). The source term $S(\theta,\phi,t,Class)$ is determined based on the statistical properties of the classes for the previous rotations at a given heliografic sector.

\section{Concluding remarks}

In the present work, we have developed an artificial neural network model to predict the short-term evolution of the solar irradiance based on near-real time observations of the solar surface magnetic field. We have shown that the total and spectral solar irradiance can be predicted up to three days with high accuracy by considering the evolution and distribution of magnetic structures on the solar disk. 

In order to reduce the dimensionality of the problem, we have employed a feature extraction algorithm to identify and classify the magnetic structures observed on the solar disk. The classification scheme is based on the empirical distribution function of the fraction of the solar disk covered by individual structures. In this way, we have considered separately the evolution of large active regions and small short-lived ephemeral regions. Additionally, we have considered the evolution of sunspots (umbrae and penumbrae).

The coefficients of the neural network are constrained by comparing the output of the model and measurements of the solar irradiance by instruments onboard of SORCE spacecraft. The generalization of the network is tested by dividing the data sets on two groups: (1) the training set; and, (2) the validation set. We have found that the model error is wavelength dependent. While the model error for 24-hour forecast in the band from 115 to 180 nm is lower than 5\%, the model error can reach 20\% in the band from 180 to 310 nm. We speculate that the difference between the performance of the network for these two bands can be due to the degradation and reduction of the accuracy of the MUV measurements. 

We have also tested the performance of the network for different forecast periods. As expected, the performance of the network reduces progressively with the increase of the forecast period, which limits significantly the maximum forecast period that we can achieve with the discussed architecture. In Sect 4.3 we discuss briefly how these limitations can be overcome by employing a solar surface magnetic flux transport model.

The real-time short-term forecast of the total and spectral solar irradiance is available at http://www.lpc2e.cnrs-orleans.fr/~soteria. The extension of the model to wavelengths above 310 nm will be available by the end of 2011. 

%

%

%

%

%
%
\bibliographystyle{spr-mp-sola}
 \bibliography{SSIversion03}  

\begin{thebibliography}{26}
\ifx \bisbn   \undefined \def \bisbn  #1{ISBN #1}\fi
\ifx \binits  \undefined \def \binits#1{#1}\fi
\ifx \bauthor  \undefined \def \bauthor#1{#1}\fi
\ifx \batitle  \undefined \def \batitle#1{#1}\fi
\ifx \bjtitle  \undefined \def \bjtitle#1{\textit{#1}}\fi
\ifx \bvolume  \undefined \def \bvolume#1{\textbf{#1}}\fi
\ifx \byear  \undefined \def \byear#1{#1}\fi
\ifx \bissue  \undefined \def \bissue#1{#1}\fi
\ifx \bfpage  \undefined \def \bfpage#1{#1}\fi
\ifx \blpage  \undefined \def \blpage #1{#1}\fi
\ifx \burl  \undefined \def \burl#1{\textsf{#1}}\fi
\ifx \href  \undefined \def \href#1#2{\textsf{#2}}\fi
\ifx \doiurl  \undefined \def
  \doiurl#1{\href{http://dx.doi.org/#1}{\textsf{#1}}}\fi
\ifx \betal  \undefined \def \betal{\textit{et al.}}\fi
\ifx \binstitute  \undefined \def \binstitute#1{#1}\fi
\ifx \bctitle  \undefined \def \bctitle#1{#1}\fi
\ifx \beditor  \undefined \def \beditor#1{#1}\fi
\ifx \bpublisher  \undefined \def \bpublisher#1{#1}\fi
\ifx \bbtitle  \undefined \def \bbtitle#1{\textit{#1}}\fi
\ifx \bedition  \undefined \def \bedition#1{#1}\fi
\ifx \bseriesno  \undefined \def \bseriesno#1{\textbf{#1}}\fi
\ifx \blocation  \undefined \def \blocation#1{#1}\fi
\ifx \bsertitle  \undefined \def \bsertitle#1{\textit{#1}}\fi
\ifx \bsnm \undefined \def \bsnm#1{#1}\fi
\ifx \bsuffix \undefined \def \bsuffix#1{#1}\fi
\ifx \bparticle \undefined \def \bparticle#1{#1}\fi
\ifx \barticle \undefined \def \barticle#1{}\fi
\ifx \botherref \undefined \def \botherref#1{}\fi
\ifx \url \undefined \def \url#1{\textsf{#1}}\fi
\ifx \bchapter \undefined \def \bchapter#1{}\fi
\ifx \bbook \undefined \def \bbook#1{}\fi
\ifx \bcomment \undefined \def \bcomment#1{#1}\fi
\ifx \oauthor \undefined \def \oauthor#1{#1}\fi
\ifx \citeauthoryear \undefined \def \citeauthoryear#1{#1}\fi
\def \endbibitem {}
\ifx \bconflocation  \undefined \def \bconflocation#1{#1} \fi

\bibitem[\protect\citeauthoryear{{Ball} \textit{et~al.}}{2011}]{ball11}
\begin{barticle}
\bauthor{\bsnm{{Ball}}, \binits{W.T.}},
\bauthor{\bsnm{{Unruh}}, \binits{Y.C.}},
\bauthor{\bsnm{{Krivova}}, \binits{N.A.}},
\bauthor{\bsnm{{Solanki}}, \binits{S.}},
\bauthor{\bsnm{{Harder}}, \binits{J.W.}}:
\byear{2011},
\batitle{{Solar irradiance variability: a six-year comparison between SORCE
  observations and the SATIRE model}}.
\bjtitle{Astronomy and Astrophysics}
\bvolume{530},
\bfpage{A71}.
doi:\doiurl{10.1051/0004-6361/201016189}.
\end{barticle}
\endbibitem

\bibitem[\protect\citeauthoryear{{Chamberlin}, {Woods}, and
  {Eparvier}}{2008}]{chamberlin08}
\begin{barticle}
\bauthor{\bsnm{{Chamberlin}}, \binits{P.C.}},
\bauthor{\bsnm{{Woods}}, \binits{T.N.}},
\bauthor{\bsnm{{Eparvier}}, \binits{F.G.}}:
\byear{2008},
\batitle{{New flare model using recent measurements of the solar ultraviolet
  irradiance}}.
\bjtitle{Adv. Space Research}
\bvolume{42},
\bfpage{912}\,--\,\blpage{916}.
doi:\doiurl{10.1016/j.asr.2007.09.009}.
\end{barticle}
\endbibitem

\bibitem[\protect\citeauthoryear{{Domingo} \textit{et~al.}}{2009}]{domingo09}
\begin{barticle}
\bauthor{\bsnm{{Domingo}}, \binits{V.}},
\bauthor{\bsnm{{Ermolli}}, \binits{I.}},
\bauthor{\bsnm{{Fox}}, \binits{P.}},
\bauthor{\bsnm{{Fr{\"o}hlich}}, \binits{C.}},
\bauthor{\bsnm{{Haberreiter}}, \binits{M.}},
\bauthor{\bsnm{{Krivova}}, \binits{N.}},
\bauthor{\bsnm{{Kopp}}, \binits{G.}},
\bauthor{\bsnm{{Schmutz}}, \binits{W.}},
\bauthor{\bsnm{{Solanki}}, \binits{S.K.}},
\bauthor{\bsnm{{Spruit}}, \binits{H.C.}},
\bauthor{\bsnm{{Unruh}}, \binits{Y.}},
\bauthor{\bsnm{{V{\"o}gler}}, \binits{A.}}:
\byear{2009},
\batitle{{Solar Surface Magnetism and Irradiance on Time Scales from Days to
  the 11-Year Cycle}}.
\bjtitle{Space Science Reviews}
\bvolume{145},
\bfpage{337}\,--\,\blpage{380}.
doi:\doiurl{10.1007/s11214-009-9562-1}.
\end{barticle}
\endbibitem

\bibitem[\protect\citeauthoryear{Elman}{1990}]{elman90}
\begin{barticle}
\bauthor{\bsnm{Elman}, \binits{J.L.}}:
\byear{1990},
\batitle{Finding structure in time}.
\bjtitle{Cognitive Science}
\bvolume{14},
\bfpage{179}\,--\,\blpage{211}.
\end{barticle}
\endbibitem

\bibitem[\protect\citeauthoryear{{Fligge} \textit{et~al.}}{1998}]{fligge98}
\begin{barticle}
\bauthor{\bsnm{{Fligge}}, \binits{M.}},
\bauthor{\bsnm{{Solanki}}, \binits{S.K.}},
\bauthor{\bsnm{{Unruh}}, \binits{Y.C.}},
\bauthor{\bsnm{{Froehlich}}, \binits{C.}},
\bauthor{\bsnm{{Wehrli}}, \binits{C.}}:
\byear{1998},
\batitle{{A model of solar total and spectral irradiance variations}}.
\bjtitle{Astronomy and Astrophysics}
\bvolume{335},
\bfpage{709}\,--\,\blpage{718}.
\end{barticle}
\endbibitem

\bibitem[\protect\citeauthoryear{Fontenla \textit{et~al.}}{2011}]{fontenla11}
\begin{barticle}
\bauthor{\bsnm{Fontenla}, \binits{J.M.}},
\bauthor{\bsnm{Harder}, \binits{J.}},
\bauthor{\bsnm{Livingston}, \binits{W.}},
\bauthor{\bsnm{Snow}, \binits{M.}},
\bauthor{\bsnm{Woods}, \binits{T.}}:
\byear{2011},
\batitle{{High-resolution solar spectral irradiance from extreme ultraviolet to
  far infrared}}.
\bjtitle{Journal of Geophysical Research (Atmospheres)}
\bvolume{116}(\bissue{D20}).
\bisbn{0148-0227}.
doi:\doiurl{10.1029/2011JD016032}.
\burl{http://dx.doi.org/10.1029/2011JD016032}.
\end{barticle}
\endbibitem

\bibitem[\protect\citeauthoryear{{Fr{\"o}hlich}}{2011}]{froehlich11}
\begin{botherref}
\oauthor{\bsnm{{Fr{\"o}hlich}}, \binits{C.}}:
2011,
{Total Solar Irradiance: What Have We Learned from the Last Three Cycles and
  the Recent Minimum?}
\textit{Space Science Reviews},
66.
doi:\doiurl{10.1007/s11214-011-9780-1}.
\end{botherref}
\endbibitem

\bibitem[\protect\citeauthoryear{Gray \textit{et~al.}}{2010}]{gray10}
\begin{barticle}
\bauthor{\bsnm{Gray}, \binits{L.J.}},
\bauthor{\bsnm{Beer}, \binits{J.}},
\bauthor{\bsnm{Geller}, \binits{M.}},
\bauthor{\bsnm{Haigh}, \binits{J.D.}},
\bauthor{\bsnm{Lockwood}, \binits{M.}},
\bauthor{\bsnm{Matthes}, \binits{K.}},
\bauthor{\bsnm{Cubasch}, \binits{U.}},
\bauthor{\bsnm{Fleitmann}, \binits{D.}},
\bauthor{\bsnm{Harrison}, \binits{G.}},
\bauthor{\bsnm{Hood}, \binits{L.}},
\bauthor{\bsnm{Luterbacher}, \binits{J.}},
\bauthor{\bsnm{Meehl}, \binits{G.A.}},
\bauthor{\bsnm{Shindell}, \binits{D.}},
\bauthor{\bparticle{van} \bsnm{Geel}, \binits{B.}},
\bauthor{\bsnm{White}, \binits{W.}}:
\byear{2010},
\batitle{Solar influences on climate}.
\bjtitle{Rev. Geophys.}
\bvolume{48}(\bissue{4}),
\bfpage{1}\,--\,\blpage{53}.
\bisbn{8755-1209}.
\burl{http://dx.doi.org/10.1029/2009RG000282}.
\end{barticle}
\endbibitem

\bibitem[\protect\citeauthoryear{{Haigh}}{2007}]{haigh07}
\begin{barticle}
\bauthor{\bsnm{{Haigh}}, \binits{J.D.}}:
\byear{2007},
\batitle{{The Sun and the Earth's Climate}}.
\bjtitle{Living Reviews in Solar Physics}
\bvolume{4},
\bfpage{2}\,--\,\blpage{65}.
\burl{http://solarphysics.livingreviews.org/Articles/lrsp-2007-2/}.
\end{barticle}
\endbibitem

\bibitem[\protect\citeauthoryear{{Haigh} \textit{et~al.}}{2010}]{haigh10}
\begin{barticle}
\bauthor{\bsnm{{Haigh}}, \binits{J.D.}},
\bauthor{\bsnm{{Winning}}, \binits{A.R.}},
\bauthor{\bsnm{{Toumi}}, \binits{R.}},
\bauthor{\bsnm{{Harder}}, \binits{J.W.}}:
\byear{2010},
\batitle{{An influence of solar spectral variations on radiative forcing of
  climate}}.
\bjtitle{Nature}
\bvolume{467},
\bfpage{696}\,--\,\blpage{699}.
doi:\doiurl{10.1038/nature09426}.
\end{barticle}
\endbibitem

\bibitem[\protect\citeauthoryear{Harder \textit{et~al.}}{2009}]{harder09}
\begin{botherref}
\oauthor{\bsnm{Harder}, \binits{J.W.}},
\oauthor{\bsnm{Fontenla}, \binits{J.M.}},
\oauthor{\bsnm{Pilewskie}, \binits{P.}},
\oauthor{\bsnm{Richard}, \binits{E.C.}},
\oauthor{\bsnm{Woods}, \binits{T.N.}}:
2009,
Trends in solar spectral irradiance variability in the visible and infrared.
\textit{Geoph. Res. Lett.}
\textbf{36}.
\url{http://dx.doi.org/10.1029/2008GL036797}.
\end{botherref}
\endbibitem

\bibitem[\protect\citeauthoryear{Harvey}{1993}]{harvey1993}
\begin{botherref}
\oauthor{\bsnm{Harvey}, \binits{K.}}:
1993,
Magnetic bipoles on the sun.
Ph.d. thesis,
University of Utrecht.
\end{botherref}
\endbibitem

\bibitem[\protect\citeauthoryear{Jiang \textit{et~al.}}{2010}]{jiang2010}
\begin{barticle}
\bauthor{\bsnm{Jiang}, \binits{J.}},
\bauthor{\bsnm{Cameron}, \binits{R.}},
\bauthor{\bsnm{Schmitt}, \binits{D.}},
\bauthor{\bsnm{Schüssler}, \binits{M.}}:
\byear{2010},
\batitle{Modeling the sun's open magnetic flux and the heliospheric current
  sheet}.
\bjtitle{The Astrophysical Journal}
\bvolume{709}(\bissue{1}),
\bfpage{301}.
\burl{http://stacks.iop.org/0004-637X/709/i=1/a=301}.
\end{barticle}
\endbibitem

\bibitem[\protect\citeauthoryear{{Krivova} and {Solanki}}{2008}]{krivova08}
\begin{barticle}
\bauthor{\bsnm{{Krivova}}, \binits{N.A.}},
\bauthor{\bsnm{{Solanki}}, \binits{S.K.}}:
\byear{2008},
\batitle{{Models of solar irradiance variations: Current status}}.
\bjtitle{Journal of Astrophysics and Astronomy}
\bvolume{29},
\bfpage{151}\,--\,\blpage{158}.
doi:\doiurl{10.1007/s12036-008-0018-x}.
\end{barticle}
\endbibitem

\bibitem[\protect\citeauthoryear{Krivova, Vieira, and
  Solanki}{2010}]{krivova10}
\begin{botherref}
\oauthor{\bsnm{Krivova}, \binits{N.A.}},
\oauthor{\bsnm{Vieira}, \binits{L.E.A.}},
\oauthor{\bsnm{Solanki}, \binits{S.K.}}:
2010,
{Reconstruction of solar spectral irradiance since the Maunder minimum}.
\textit{Journal of Geophysical Research (Space Physics)}
\textbf{115}(A12).
0148-0227.
\url{http://dx.doi.org/10.1029/2010JA015431}.
\end{botherref}
\endbibitem

\bibitem[\protect\citeauthoryear{Krivova \textit{et~al.}}{2003}]{krivova2003}
\begin{barticle}
\bauthor{\bsnm{Krivova}, \binits{N.A.}},
\bauthor{\bsnm{Solanki}, \binits{S.K.}},
\bauthor{\bsnm{Fligge}, \binits{M.}},
\bauthor{\bsnm{Unruh}, \binits{Y.C.}}:
\byear{2003},
\batitle{Reconstruction of solar irradiance variations in cycle 23: Is solar
  surface magnetism the cause?}
\bjtitle{\aap}
\bvolume{399}(\bissue{1}),
\bfpage{L1}\,--\,\blpage{L4}.
\end{barticle}
\endbibitem

\bibitem[\protect\citeauthoryear{{Lean}}{2000}]{lean00b}
\begin{barticle}
\bauthor{\bsnm{{Lean}}, \binits{J.L.}}:
\byear{2000},
\batitle{{Evolution of the Sun's spectral irradiance since the Maunder
  Minimum}}.
\bjtitle{Geoph. Res. Lett.}
\bvolume{27},
\bfpage{2425}\,--\,\blpage{2428}.
doi:\doiurl{10.1029/2000GL000043}.
\end{barticle}
\endbibitem

\bibitem[\protect\citeauthoryear{{Lean}}{2005}]{lean05}
\begin{barticle}
\bauthor{\bsnm{{Lean}}, \binits{J.L.}}:
\byear{2005},
\batitle{{Living with a Variable Sun}}.
\bjtitle{Physics Today}
\bvolume{58},
\bfpage{32}\,--\,\blpage{38}.
doi:\doiurl{10.1063/1.1996472}.
\end{barticle}
\endbibitem

\bibitem[\protect\citeauthoryear{{Lean} \textit{et~al.}}{2011}]{lean11}
\begin{barticle}
\bauthor{\bsnm{{Lean}}, \binits{J.L.}},
\bauthor{\bsnm{{Woods}}, \binits{T.N.}},
\bauthor{\bsnm{{Eparvier}}, \binits{F.G.}},
\bauthor{\bsnm{{Meier}}, \binits{R.R.}},
\bauthor{\bsnm{{Strickland}}, \binits{D.J.}},
\bauthor{\bsnm{{Correira}}, \binits{J.T.}},
\bauthor{\bsnm{{Evans}}, \binits{J.S.}}:
\byear{2011},
\batitle{{Solar extreme ultraviolet irradiance: Present, past, and future}}.
\bjtitle{Journal of Geophysical Research (Space Physics)}
\bvolume{116},
\bfpage{A01102}.
doi:\doiurl{10.1029/2010JA015901}.
\end{barticle}
\endbibitem

\bibitem[\protect\citeauthoryear{{Lilensten}
  \textit{et~al.}}{2008}]{lilensten08}
\begin{barticle}
\bauthor{\bsnm{{Lilensten}}, \binits{J.}},
\bauthor{\bsnm{{Dudok de Wit}}, \binits{T.}},
\bauthor{\bsnm{{Kretzschmar}}, \binits{M.}},
\bauthor{\bsnm{{Amblard}}, \binits{P.-O.}},
\bauthor{\bsnm{{Moussaoui}}, \binits{S.}},
\bauthor{\bsnm{{Aboudarham}}, \binits{J.}},
\bauthor{\bsnm{{Auch{\`e}re}}, \binits{F.}}:
\byear{2008},
\batitle{{Review on the solar spectral variability in the EUV for space weather
  purposes}}.
\bjtitle{Annales Geophysicae}
\bvolume{26},
\bfpage{269}\,--\,\blpage{279}.
\end{barticle}
\endbibitem

\bibitem[\protect\citeauthoryear{MacKay}{1992}]{MacKay1992}
\begin{barticle}
\bauthor{\bsnm{MacKay}}:
\byear{1992},
\batitle{Mackay1992}.
\bjtitle{Neural Computation}
\bvolume{4},
\bfpage{415}\,--\,\blpage{447}.
\end{barticle}
\endbibitem

\bibitem[\protect\citeauthoryear{Rottman}{2005}]{rottman2005}
\begin{barticle}
\bauthor{\bsnm{Rottman}, \binits{G.}}:
\byear{2005},
\batitle{The sorce mission}.
\bjtitle{Solar Physics}
\bvolume{230},
\bfpage{7}\,--\,\blpage{25}.
\bcomment{10.1007/s11207-005-8112-6}.
\burl{http://dx.doi.org/10.1007/s11207-005-8112-6}.
\end{barticle}
\endbibitem

\bibitem[\protect\citeauthoryear{Schou \textit{et~al.}}{2011}]{schou2011}
\begin{botherref}
\oauthor{\bsnm{Schou}, \binits{J.}},
\oauthor{\bsnm{Scherrer}, \binits{P.}},
\oauthor{\bsnm{Bush}, \binits{R.}},
\oauthor{\bsnm{Wachter}, \binits{R.}},
\oauthor{\bsnm{Couvidat}, \binits{S.}},
\oauthor{\bsnm{Rabello-Soares}, \binits{M.}},
\oauthor{\bsnm{Bogart}, \binits{R.}},
\oauthor{\bsnm{Hoeksema}, \binits{J.}},
\oauthor{\bsnm{Liu}, \binits{Y.}},
\oauthor{\bsnm{Duvall}, \binits{T.}},
\oauthor{\bsnm{Akin}, \binits{D.}},
\oauthor{\bsnm{Allard}, \binits{B.}},
\oauthor{\bsnm{Miles}, \binits{J.}},
\oauthor{\bsnm{Rairden}, \binits{R.}},
\oauthor{\bsnm{Shine}, \binits{R.}},
\oauthor{\bsnm{Tarbell}, \binits{T.}},
\oauthor{\bsnm{Title}, \binits{A.}},
\oauthor{\bsnm{Wolfson}, \binits{C.}},
\oauthor{\bsnm{Elmore}, \binits{D.}},
\oauthor{\bsnm{Norton}, \binits{A.}},
\oauthor{\bsnm{Tomczyk}, \binits{S.}}:
2011,
Design and ground calibration of the helioseismic and magnetic imager (hmi)
  instrument on the solar dynamics observatory(sdo).
\textit{Solar Physics},
1\,--\,31.
10.1007/s11207-011-9842-2.
\url{http://dx.doi.org/10.1007/s11207-011-9842-2}.
\end{botherref}
\endbibitem

\bibitem[\protect\citeauthoryear{Shapiro \textit{et~al.}}{2011}]{shapiro2011}
\begin{barticle}
\bauthor{\bsnm{Shapiro}, \binits{A.I.}},
\bauthor{\bsnm{Schmutz}, \binits{W.}},
\bauthor{\bsnm{Rozanov}, \binits{E.}},
\bauthor{\bsnm{Schoell}, \binits{M.}},
\bauthor{\bsnm{Haberreiter}, \binits{M.}},
\bauthor{\bsnm{Shapiro}, \binits{A.V.}},
\bauthor{\bsnm{Nyeki}, \binits{S.}}:
\byear{2011},
\batitle{A new approach to the long-term reconstruction of the solar irradiance
  leads to large historical solar forcing}.
\bjtitle{\aap}
\bvolume{529},
\bfpage{A67}.
doi:\doiurl{10.1051/0004-6361/201016173}.
\burl{http://dx.doi.org/10.1051/0004-6361/201016173}.
\end{barticle}
\endbibitem

\bibitem[\protect\citeauthoryear{{Tobiska} and {Bouwer}}{2006}]{tobiska06b}
\begin{barticle}
\bauthor{\bsnm{{Tobiska}}, \binits{W.K.}},
\bauthor{\bsnm{{Bouwer}}, \binits{S.D.}}:
\byear{2006},
\batitle{{New developments in SOLAR2000 for space research and operations}}.
\bjtitle{Adv. Space Research}
\bvolume{37},
\bfpage{347}\,--\,\blpage{358}.
doi:\doiurl{10.1016/j.asr.2005.08.015}.
\end{barticle}
\endbibitem

\bibitem[\protect\citeauthoryear{Vieira \textit{et~al.}}{2011}]{vieira2011}
\begin{barticle}
\bauthor{\bsnm{Vieira}, \binits{L.E.A.}},
\bauthor{\bsnm{Solanki}, \binits{S.K.}},
\bauthor{\bsnm{Krivova}, \binits{N.A.}},
\bauthor{\bsnm{Usoskin}, \binits{I.}}:
\byear{2011},
\batitle{Evolution of the solar irradiance during the holocene}.
\bjtitle{\aap}
\bvolume{531},
\bfpage{A6}.
doi:\doiurl{10.1051/0004-6361/201015843}.
\burl{http://dx.doi.org/10.1051/0004-6361/201015843}.
\end{barticle}
\endbibitem

\end{thebibliography}

%
%
%
%

\end{document}